\documentclass[pre,twocolumn,superscriptaddress,showpacs]{revtex4-1}

\usepackage{graphicx}
\usepackage{amsmath}
\usepackage{latexsym}
\usepackage{comment}
\usepackage{bm}
\usepackage{url}
\usepackage{hyperref}
\hypersetup{colorlinks,urlcolor=blue}

\begin{document}
\title{Curvature driven motion of a bubble in a toroidal Hele-Shaw cell}
\date{\today}
\author{A. Mughal}
\affiliation{Department of Mathematics, Aberystwyth University, Aberystwyth, Ceredigion SY23 3BZ, U.K.}
\affiliation{Theoretische Physik, Friedrich-Alexander-Universit\"{a}t Erlangen-N\"{u}rnberg - Staudtstr. 7, 91058 Erlangen, Germany}
\author{S. J. Cox}
\affiliation{Department of Mathematics, Aberystwyth University, Aberystwyth, Ceredigion SY23 3BZ, U.K.}
\author{G. E. Schr\"{o}der-Turk}
\affiliation{Murdoch University, School of Engineering and IT, Murdoch, Perth, WA6162, Australia}
\affiliation{Theoretische Physik, Friedrich-Alexander-Universit\"{a}t Erlangen-N\"{u}rnberg - Staudtstr. 7, 91058 Erlangen, Germany}

\begin{abstract}
We investigate the equilibrium properties of a single area-minimising bubble trapped between two narrowly-separated parallel curved plates. We begin with the case of a bubble trapped between concentric spherical plates. We develop a model which shows that the surface energy of the bubble is lower when confined between spherical plates than between flat plates. We confirm our findings by comparing against Surface Evolver simulations. We then derive a simple model for a bubble between arbitrarily curved parallel plates. The energy is found to be higher when the local Gaussian curvature of the plates is negative and lower when the curvature is positive. To check the validity of the model we consider a bubble trapped between concentric tori. In the toroidal case we find that the sensitivity of the bubble's energy to the local curvature acts as a geometric potential capable of driving bubbles from regions with negative to positive curvature.
\end{abstract}

\maketitle

\section{Introduction}
The \emph{isoperimetric problem} has fascinated mathematicians and physicists for centuries \cite{blaasjo2005isoperimetric}. One statement of the problem is to find the surface of least possible area that encloses a given volume. Due to surface tension the free energy of a soap bubble depends directly on its surface area. Soap bubbles minimise their area for the volume they enclose and therefore serve as a rich playground for isoperimetric problems \cite{weaire1997kelvin}. 

\begin{figure}
\begin{center}
\includegraphics[width=1.0\columnwidth ]{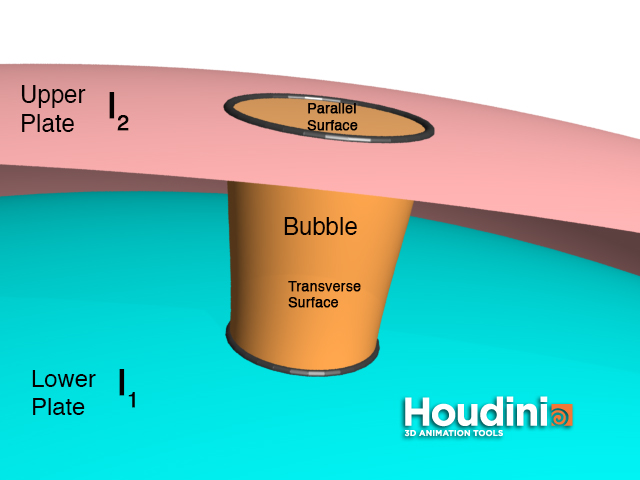}
\caption{A bubble trapped between parallel concentric spherical plates (an upper plate $I_2$ and lower plate $I_1$). The parallel surfaces of the bubble lie on the bounding plates while the transverse surface (which spans the gap) touches them at $90^o$. Only the transverse surface contributes to the surface energy of the bubble (see below). }
\label{bubblespheres}
\end{center}
\end{figure}

In two-dimensions (2D) the least perimeter way of enclosing an infinite number of equal area cells is given by the honeycomb structure \cite{hales2001honeycomb}. Such hexagonal arrangements are frequently encountered in the study of 2D foams. In the laboratory the situation can be approximated by trapping a single layer of equal-sized (monodisperse) bubbles in a so-called \emph{Hele-Shaw cell}. The cell consists of two narrowly-spaced glass plates which are (i) flat, and (ii) parallel to each other \cite{stevenson2012foam, drenckhan2010monodisperse}. The resulting optimal (least area) \emph{quasi-2D foam} is comprised of a network of hexagons, in which the liquid films are perpendicular to the glass plates \cite{huerre2014bubbles}. 

The usefulness of the Hele-Shaw arrangement goes far beyond the study of such highly-regular arrangements. Indeed it can been used to generate a wide class of disordered polydisperse quasi-2D foams. The structure and dynamics of such foams are now rather well understood \cite{cox2006shear, cox2008structure} and they have been used to inform models of three-dimensional (3D) foams. Given the influence of the Hele-Shaw cell in the study of 2D foams, it is interesting to consider the properties of a single bubble (or a foam) between narrowly separated plates in which one of the two above conditions is relaxed. 

For example, consider an ordered monodisperse foam between flat \emph{non-parallel} plates. The changing plate separation imposes a systematic variation in the \emph{apparent} area of the bubbles (i.e. the area of the bubbles as observed from above or, equivalently, projected onto one of the two plates). As a result this quasi-2D foam can be made to closely approximate various conformal maps of the undeformed hexagonal structure \cite{drenckhan2004demonstration, Mancini:2005we, Mancini:2005th, mughal2009curvature}. Another important example is the trapping of a foam between concentric hemispheres \cite{roth2012coarsening, senden}; in this case, the curvature of the plates leads to a modification of the famous von Neumann's law for the diffusion-driven coarsening of a dry 2D foam \cite{roth2012coarsening, Peczak:1993ih,  ryan2016curvature}.

There is already a considerable literature devoted to investigating the necessarily-defective crystalline phases of various soft matter systems (e.g. liquid crystals \cite{shin2008topological}, colloids \cite{bausch2003grain} and charged particles \cite{altschuler2005global}) on curved surfaces. In a flat Hele-Shaw cell we may expect that the ground-state of a monodisperse ordered 2D foam is given by a regular packing of hexagons, since this is the minimal perimeter arrangement on the plane. However, in a curved cell the symmetries of the hexagonal space group do not apply \cite{vitelli2006crystallography}. As in other similar systems, we expect this conflict to be resolved by the appearance of \emph{topological defects} such as disclinations and dislocations \cite{bowick2009two}, as observed in simulations of strictly 2D foams on the surface of a sphere \cite{cox2010minimal}. The precise number and arrangement of such defects will depend on the topology and curvature of the substrate. 

Here we show that, in addition to these topological constraints, in a \emph{curved} Hele-Shaw cell the local curvature of the plates results in a \emph{geometric potential} acting on individual bubbles. 

We consider cells where the bounding plates are curved but parallel to each other. By parallel we mean that the two plates are parallel surfaces of each other i.e. surfaces with a constant point-wise distance between them. Thus if each point ${\bf p}$ on the first surface $I_1$ is translated along the surface normal ${\bf n}$ by a distance $d$ we obtain the corresponding point on the second bounding plate \cite{do1976differential} (see Fig \ref{bubblespheres}),
\[
I_2=\{{\bf p}+d\cdot {\bf n}({\bf p}) | {\bf p} \in I_1 \}.
\] 
Here we restrict the distance $d$ to be smaller than the absolute value of the local radius of curvature; this avoids any potential singularities and ensures that the second bounding plate is a regular surface. Note that this definition obeys reciprocity: if surface $I_2$ is parallel to $I_1$, then $I_1$ is parallel to $I_2$.

Provided the gap $d$ between the plates is small, we find that bubbles in regions with a positive Gaussian curvature have a lower surface energy than those in regions with negative Gaussian curvature. This difference can be interpreted as a geometric potential that can, in principle, drive bubbles from regions with minimum (negative) to maximum (positive) Gaussian curvature. 

We anticipate that this result will be of interest in understanding not only the forces acting on single bubbles but also on clusters of bubbles and foams. Areas of applications may include microfluidics using assemblies of bubbles \cite{drenckhan2005rheology}, the adhesion and motion of single cell organisms on curved surfaces \cite{das2008adhesion} and the motion of bubbles in porous media \cite{cox2004theory}. We expect this work to contribute to related problems concerning the role of various bounding surfaces in determining bubble morphology (and dynamics), famous examples include the area-minimising Rayleigh undulation instability for a cylinder and the catenoid minimal surface between two rings. Other related problems include the study of polymers confined between curved surfaces \cite{yaman1997polymers}. 

The paper is organised as follows. In section~\ref{sec:model} we describe the surface energy of a single bubble between two surfaces. In section~\ref{sec:theorem} we develop a model to describe the energy of a bubble between two narrowly separated curved plates. We then proceed in section~\ref{sec:sims} to compare our model with numerical simulations.  We pay special attention to the case of a bubble between toroidal surfaces, since the toroidal cell includes regions of positive and negative Gaussian curvature. In section~\ref{sec:pert} we consider how this geometric potential is modified if the bounding plates are not strictly parallel. Finally, we finish with section~\ref{sec:disc} which includes a short summary and discussion of the main results.

\section{The Model}
\label{sec:model}

A bubble consists of a liquid interface with a surface tension $\gamma$ enclosing a volume of gas $V$. The total surface energy of the bubble is given by,
\[
E=\gamma A
\]
where $\gamma$ is assumed to be constant and $A$ is the surface area of the bubble.

We assume that on two sides the bubble is bounded by solid surfaces. The surfaces are not necessarily flat but are smooth and frictionless. This allows the bubble interface to slide along the walls and relax to equilibrium.

Plateau's laws, a consequence of surface energy minimisation, dictate that the interface meets the confining wall at right angles (normal incidence). The surface energy of the bubble therefore depends entirely on the surface area of the (transverse) film between the walls.

\section{Theory}
\label{sec:theorem}

We begin by considering a related problem: that of a \emph{strictly} $2D$ bubble on a sphere. We then derive an approximate model for a quasi-2D bubble between two narrowly separated spherical plates. Finally, we develop a simple (leading order) expression for the surface energy of a bubble between arbitrarily curved parallel plates. 

\subsection{2D bubble on a sphere}

\begin{figure}
\begin{center}
\includegraphics[width=0.5\columnwidth ]{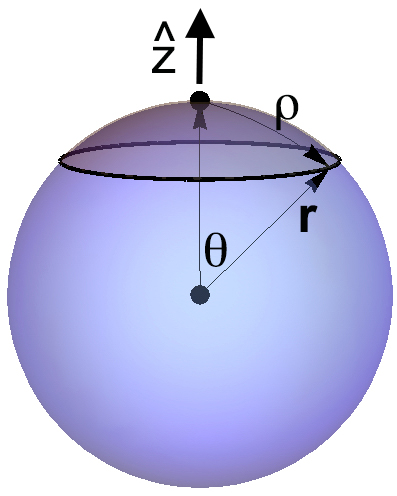}
\caption{A geodesic circle of radius $\rho$ on a sphere of radius $r$. Here $\hat{\bf{z}}$ is the unit vector in the normal direction. The geodesic circle (shown in black) has area $S(r)$ and circumference $C(r)$.}
\label{spherepatch}
\end{center}
\end{figure}

For a strictly 2D bubble of area $S$ the energy is given by $E=\gamma P$, where $P$ is the perimeter of the bubble.

The minimal energy solution for a strictly 2D bubble on a sphere of radius $r$ can be surmised from the related isoperimetric problem on a sphere. The problem is to find the shape of a given area, confined to the surface of a sphere, that has the least perimeter; the solution is known to be given by a {\em geodesic} circle \cite{canete2008isoperimetric}. Thus, as shown in Fig \ref{spherepatch}, the minimal energy solution for a strictly 2D bubble on a sphere is a spherical cap with geodesic radius $\rho$, area
\begin{eqnarray}
S(r)
&=&
\int_{0}^{2\pi}
\int_{0}^{\theta=\rho/r}
r^2\sin \theta' \textrm{d}\theta' \textrm{d}\phi'
\nonumber
\\
&=&
2\pi r^2
\left(
1-
\cos\left(\frac{\rho}{r} \right)
\right)
\label{eq:circarea0}
\end{eqnarray}
and circumference
\begin{eqnarray}
C(r)
&=&
\int_{0}^{2\pi}
r\sin \theta \textrm{d}\phi'
\nonumber
\\
&=&
2\pi r \sin \left(\frac{\rho}{r} \right),
\label{eq:circlen}
\end{eqnarray}
again using $\theta = \rho/r$.
Combining Eq. ($\!\!$~\ref{eq:circarea0}) and Eq. ($\!\!$~\ref{eq:circlen}) shows that the surface energy of a 2D bubble in terms of its area is
\begin{equation}
E^{2D}=\gamma C(r)
=
\gamma \;
2\pi r 
\sqrt{
1
-
\left(1-\frac{S}{2\pi r^2}\right)^2
}.
\end{equation}
Note that in the limit that the sphere has an infinite radius of curvature then the isoperimetric problem reduces to that of finding the minimal perimeter enclosure on a flat plane. In this case the solution is known to be given by a circle of area $S=\pi \rho^2$, thus in the limit $r\rightarrow \infty$ we find $E^{2D}\rightarrow \gamma 2\pi \rho$.
 
\subsection{Quasi-2D bubble between parallel spherical plates}

\begin{figure}
\begin{center}
\includegraphics[width=1.0\columnwidth ]{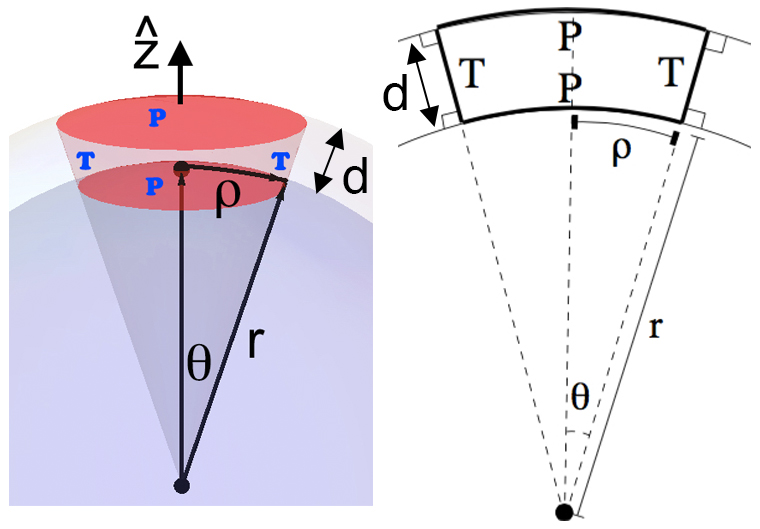}
\caption{Left: a section of a spherical cone (the red region). The films that coincide with the spherical plates (labelled P) are shown in dark red while the transverse films which are perpendicular to the spheres (labelled T) are shown in light red. Right: A schematic diagram of a spherical cone between two spherical plates of radius $r$ and $r+d$, where $d$ is the gap width.}
\label{twospheres}
\end{center}
\end{figure}

Consider a small bubble between two concentric spheres (i.e. a spherical annulus), separated by a small distance $d$, as illustrated in Fig \ref{bubblespheres}. The only contribution to the surface energy is due to the area of the transverse film which meets the bounding plates at right angles.

Provided that the gap is small then for a bubble confined between two flat plates the stable solution is known to be a cylinder. Above a critical separation $d_{crit}=(\pi V)^{1/3}$ this simple solution becomes unstable \cite{Cox:2002fz}. 

On the other hand, if the gap is small and the bounding plates are curved, then again the solution is no longer a cylinder: it is instead some other shape of constant mean curvature. So the transverse film cannot be assumed to be perpendicular everywhere to the bounding plates (that is in the direction of the surface normal to the bounding plates) although it must still meet the bounding surfaces at a right angle.

However, if both the curvature of the plates and the plate separation are sufficiently small then these deviations can be neglected. Thus a simple model of a bubble between spherical plates which ensures the condition of normal incidence is to regard it as a section of a spherical cone, as shown in Fig \ref{twospheres}. The sides of the cone represent the transverse surfaces (labelled $T$) and the spherical caps (labelled $P$) are the parallel surfaces.

For a spherical cone, with cone angle $\theta=\rho/r$, the surface area of the transverse film is
\begin{eqnarray}
A_T
&=&
\int_{0}^{2\pi}
\int_{r}^{r+d}
r'\sin\theta
\textrm{d}r'
\textrm{d}\phi'
\nonumber
\\
&=&
\pi r^2 
\sigma_2
\sin
\left(
\frac{\rho}{r}
\right),
\label{eq:areacone}
\end{eqnarray}
while the volume is given by
\begin{eqnarray}
V
&=&
\int_{r}^{r+d}
\int_{0}^{\theta=\rho/r}
\int_{0}^{2\pi}
r'^2 \sin\theta'
\textrm{d}r'
\textrm{d}\theta'
\textrm{d}\phi'
\nonumber
\\
&=&
\frac{2\pi r^3}{3}
\sigma_3
\left(
1-
\cos\left( \frac{\rho}{r}\right) 
\right),
\label{eq:volcone}
\end{eqnarray}
where 
\[
\sigma_n
=
\sigma_n{\left(\frac{d}{r}\right)}
=
\left(
1
+
\left(
\frac{d}{r}
\right)
\right)^n
-1
.
\]
Thus although a spherical cone is not a surface of constant mean curvature, it can serve as a useful approximation to the surface of a bubble between concentric spheres. The surface energy of a spherical cone is given by
\begin{equation}
E
=
\gamma A_T
=
\gamma 
\pi r^2 
\sigma_2
\sin
\left(
\frac{\rho}{r}
\right),
\end{equation}
which can be combined with Eq. ($\!\!$~\ref{eq:volcone}) to give
\begin{equation}
E
=
\gamma \;
\pi r^2
\sigma_2
\sqrt{
1
-
\left(
1
-
\frac{3V}{2\pi\sigma_3r^3}
\right)^2.
\label{eq:scenergy}
}
\end{equation}

In the limit of a bubble with a small cone angle, and provided that the bubble is trapped between narrowly separated plates, i.e.
\begin{equation}
\theta=\frac{\rho}{r}\ll 1 \;\;\;\;\; \textrm{and}  \;\;\;\;\; \frac{d}{r}\ll 1,
\label{eq:spscale}
\end{equation}
then Eq. ($\!\!$~\ref{eq:areacone}), Eq. ($\!\!$~\ref{eq:volcone}) and Eq. ($\!\!$~\ref{eq:scenergy}) reduce to the case of a cylindrical bubble with area $A_T\rightarrow2\pi\rho d$, volume $V\rightarrow\pi \rho^2 d$ and surface energy 
\begin{equation}
E\rightarrow E_0 =\gamma \; 2\pi \rho d,
\label{eq:fenergy}
\end{equation} 
where $E_0$ is the surface energy of a cylindrical bubble trapped between flat parallel plates.

The two quantities from Eq. ($\!\!$~\ref{eq:spscale}), i.e. the cone angle and the relative plate separation, can more generally be expressed in terms of the Gaussian curvature $G$:
\begin{equation}
\theta=\rho\sqrt{|G|}  \;\;\;\;\; \textrm{and}  \;\;\;\;\;d^*= d\sqrt{|G|}.
\label{eq:scale}
\end{equation}
We also consider a dimensionless form of the bubble volume by normalising the volume with the plate separation:
\begin{equation}
V^*=\frac{V}{d^3}.
\label{eq:vscale}
\end{equation}
These three dimensionless quantities characterise the properties of a bubble between curved plates. They are used below to specify our simulation parameters.

Provided that both $\theta\ll 1$ and $d^{*} \ll 1$, we can usefully model the bubble as a cylinder. This leads to a second even simpler model as we describe immediately below. 

\subsection{Quasi-2D bubble between curved plates}

In the case of a bubble between spherical plates, the spherical cone model can provide an approximation to the true surface of constant mean curvature. The model obeys the condition of normal incidence. However, we can construct curved Hele-Shaw geometries in which the bounding plates do not have rotational symmetry, for example a bubble between parallel saddle shaped plates. In such cases we resort to an even simpler approximation: we consider that both the cone angle $\theta$ and the relative plate separation $d^{*}$ are sufficiently small that a bubble between parallel curved plates can be modelled as a cylinder. Although this crude model does not obey the condition for normal incidence (since a cylindrical surface can be perpendicular to the two bounding plates \emph{only} if they are both locally flat), it nevertheless provides a valuable quantitative insight into the role of Gaussian curvature in determining the energy of a bubble. 

A simple scaling for bubbles confined between parallel plates in the presence of curvature can be derived from the \emph{Bertrand-Diquet-Puiseux theorem}. The theorem relates the circumference (or area) of a circle in flat space to that of a geodesic circle on a curved surface \cite{spivak1981comprehensive}.

\begin{figure}
\begin{center}
\includegraphics[width=1.0\columnwidth ]{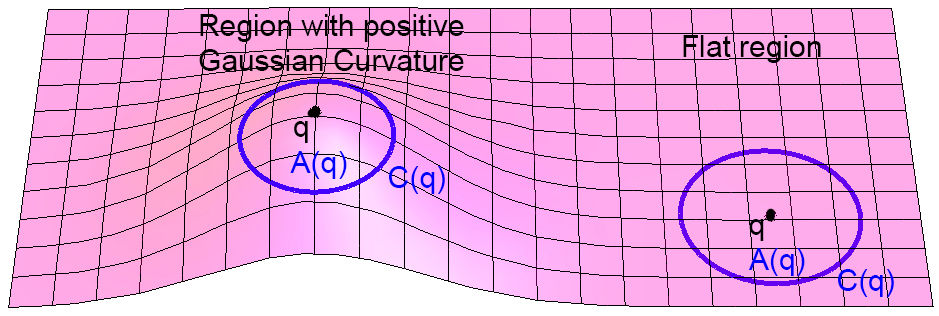}
\caption{The two blue circles represent geodesic circles about a point $q$ (black dot), with both circles having the same geodesic radius $\rho$. In a region of positive Gaussian curvature the circumference $C(q)$ of the geodesic circle is smaller than the circumference of the circle in flat space. Conversely the area $A(q)$ of the geodesic circle is slightly smaller in a region of positive Gaussian curvature.}
\label{circles}
\end{center}
\end{figure}

We define a geodesic circle of radius $\rho$ centred at $q$ as the set of all points whose geodesic distance from $q$ is equal to $\rho$, as illustrated by the blue (contours) circles in Fig \ref{circles} (note the contours of a geodesic circle are not necessarily geodesics of the surface within which the circle is embedded). 

Let $G(q)$ be the local Gaussian curvature, then the circumference of the geodesic circle is given by \cite{spivak1981comprehensive}
\begin{equation}
C(q)=2\pi \rho - \frac{\pi \rho^3 }{3}G(q)+ \mathcal{O}(\rho^5)
\label{eq:C1theorem}
\end{equation}
and the area $A(q)$ of the geodesic circle is,
\begin{equation}
A(q)=\pi \rho^2 - \frac{\pi \rho^4 }{12}G(q) +\mathcal{O}(\rho^6).
\label{eq:A1theorem}
\end{equation}
For both expressions these are the first two terms of an alternating series; the next higher order terms can be found in the appendix. Thus according to Eq. ($\!\!$~\ref{eq:C1theorem}) and Eq. ($\!\!$~\ref{eq:A1theorem}), on a surface of positive curvature the circumference (or area) of a geodesic circle is slightly ``too small'' while on a negatively curved surface it is slightly ``too large''. 

For simplicity, we shall restrict ourselves to using the first two terms in determining the circumference and area of a geodesic circle, in which case the second terms in both Eq. ($\!\!$~\ref{eq:C1theorem}) and Eq. ($\!\!$~\ref{eq:A1theorem}) can be regarded as corrections to the leading order terms. We can therefore determine the condition under which these terms remain smaller than the leading terms. Of the two, Eq. ($\!\!$~\ref{eq:C1theorem}) provides a more stringent bound, hence we require,
\[
2\pi \rho \geq \frac{\pi \rho^3}{3}|G(q)|
\]
which gives
\begin{equation}
\rho \leq \sqrt{\frac{6}{|G(q)|}}=\sqrt{6|r_1||r_2|}
\label{eq:condition}
\end{equation}
where we have written the Gaussian curvature in terms of the principle radii of curvature $r_1$ and $r_2$ in order to make the comparison with the geodesic radius $\rho$ more apparent. 

\begin{figure}
\begin{center}
\includegraphics[width=1.0\columnwidth ]{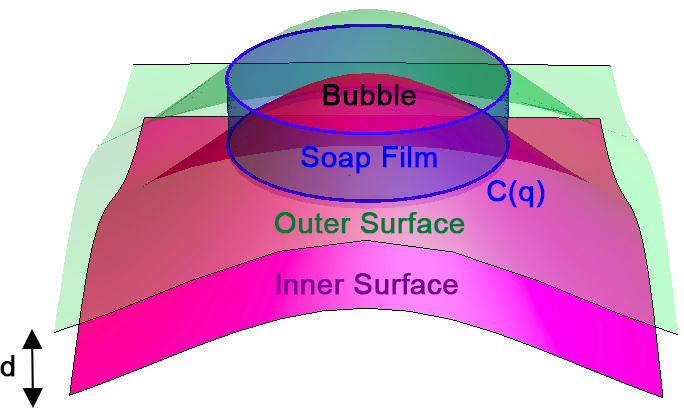}
\caption{A cylindrical bubble between two parallel curved plates separated by a distance $d$. The surface energy of the bubble depends only on the area of the transverse surface (blue shaded region) between the bounding plates.} 
\label{bubbles}
\end{center}
\end{figure}

We are now ready to consider a bubble trapped between two bounding plates that are separated by a distance $d$. The plates are assumed to be parallel and to posses some local Gaussian curvature $G(q)$, as shown in Fig \ref{bubbles}. The Gaussian curvature for the two plates may not necessarily be the same, but if the separation between the plates is sufficiently small this difference can be considered negligible. Thus from Eq. ($\!\!$~\ref{eq:C1theorem}) the surface area of the bubble (i.e. the area of the transverse interface) is approximately 
\[
A_T \approx d \; C(q) \approx d\left(2\pi \rho - \frac{\pi \rho^3}{3}G(q)  \right).
\]
The first term gives the surface area of a bubble between flat plates. Assuming a cylindrical bubble, the volume is given by $V=\rho^2 \pi d$ which can be rearranged to yield
\begin{equation}
\rho\approx\sqrt{V/\pi d}.
\label{eq:grad}
\end{equation}
This is necessarily a coarse approximation but it is one which, for the small bubbles considered here, is nevertheless broadly in agreement with the numerical results presented below. A more systematic expansion for the volume of a bubble between curved plates can be made by employing Steiner's formula \cite{turk2011}.

Thus using Eq. ($\!\!$~\ref{eq:grad}), the energy of the bubble is approximately
\begin{equation}
E=\gamma A_T=E_0- E_1\;G(q),
\label{eq:approxE}
\end{equation}
where
\begin{equation}
E_0=2\gamma \sqrt{\pi V d} \;\;\;\;\;\;  \textrm{and} \;\;\;\;\;\; E_1=\gamma\frac{1}{3\sqrt{\pi}}\sqrt{\frac{V^3}{d}}.
\label{eq:Ecoefs}
\end{equation}
Clearly in the limit that the plate curvature vanishes (i.e. $G \rightarrow 0$) we obtain $E \rightarrow E_0$. However, if the local Gaussian curvature of the bounding plates is positive then surface area (and consequently the surface energy) of the bubble is lowered. The opposite is true in a region of negative Gaussian curvature.

This simple model presented is a leading order correction to the case of a bubble trapped between narrowly-separated flat plates. We anticipate that a more accurate (and more complicated model) would also have to account for the fact that the shape of the bubble between arbitrarily curved substrates is not cylindrical, but in fact asymmetric. A further improvement would be to adjust the shape of the bubble so that it obeys the condition of normal incidence at both substrates, however this would also introduce the related problem of maintaining constant mean curvature over the entire bubble surface (or a leading approximation to this).

\begin{figure*}
\begin{center}
\includegraphics[width=2.0\columnwidth ]{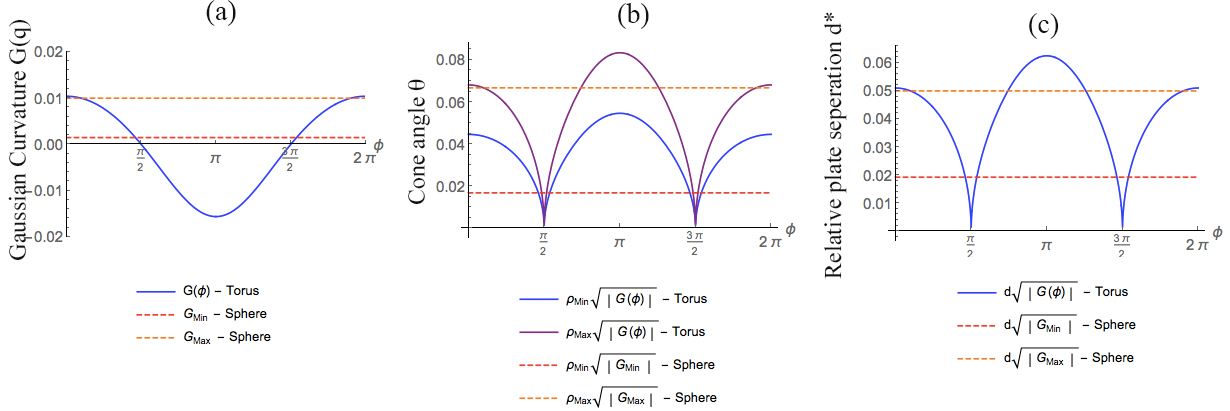}
\caption{(a) A plot of the Gaussian curvature of the inner bounding plate used to constrain the bubble. The orange and red dotted lines are the maximum and minimum values (respectively) in the spherical case. These values are compared against the toroidal case (blue solid line) where the Gaussian curvature over the inner surface is a function of the angular position $\phi$. (b) A plot of the cone angle $\theta=\rho \sqrt{|G(q)|}$.  Dotted lines indicate the maximum and minimum values obtained in simulations for bubbles between concentric spheres. Also shown for comparison are the maximum (purple) and minimum (blue) range for bubbles between concentric tori. (c) A plot of the relative plate separation  $d^{*}=d \sqrt{|G(q)|}$. Again dotted lines are maximum and minimum values in simulations for concentric spheres while the blue (solid) curve is the range of values explored for bubbles between concentric tori. }
\label{gcurvature}
\end{center}
\end{figure*}

\section{Simulations}
\label{sec:sims}

Simulations were conducted using the Surface Evolver package \cite{brakke1992surface}, which is an interactive \emph{finite element} program for the study of interfaces shaped by surface tension. Starting with a coarse mesh we generate a bubble between two plates, which are defined as constraints on the ends of the bubble. The energy of the bubble is minimised by applying gradient descent 
while repeatedly refining the mesh to improve accuracy. 

In the following we describe Surface Evolver simulations of bubbles between concentric spheres and bubbles between concentric tori. The results are compared with the models developed above. Furthermore, in order to show that the dimensions used in the two problems are of a similar magnitude, we plot the dimensionless quantities given by Eq. ($\!\!$~\ref{eq:scale}) on the same graph, see Fig \ref{gcurvature} and below for further details.

\subsection{Concentric Spheres}

We begin with the simplest possible problem involving curved plates: a bubble trapped between two concentric spheres. Despite the fact that the Gaussian curvature of the bounding plates is constant everywhere, this problem is useful as it demonstrates that the energy of a bubble is lower in a region of positive curvature.

\subsubsection{Plate Geometry}

The inner bounding plate is described by the equation $\Phi(x,y,z)=r$ and the outer plate is given by $\Phi(x,y,z)=r+d$ where $d$ is the gap width and $\Phi=\sqrt{x^2 + y^2 + z^2}$. The Gaussian curvature of the sphere is constant and in the case of the inner plate is given by $G(r)=\frac{1}{r^2}$.

\subsubsection{Simulation Parameters}

In our simulations we explore a range of bubble volumes $V$, beginning with $V=0.3$ and increasing in steps of $0.1$ up to $V=0.7$, while keeping the plate separation at a constant $d=0.5$. In terms of the dimensionless volume $V^*=V/d^3$ these values correspond to the range $2.4\leq V^*\leq5.6$.

We vary the radius of the inner shell in the range  $10 \leq r\leq 26$. The corresponding range of the Gaussian curvature is shown in Fig \ref{gcurvature}a; the minimum value is indicated by the red dotted line and the maximum value is given by the orange dotted line. 

From Eq. ($\!\!$~\ref{eq:grad}) we estimate the geodesic radius of the bubble trapped between the concentric shells to lie between the limits,
\[
\rho_{min}=0.4370 \leq \rho \leq \rho_{max}=0.6676.
\]
Using this, the range of the cone angle $\theta=\rho\sqrt{|G(q)|}$  for the bubbles in the simulation can be computed, see Fig \ref{gcurvature}b. Here $G_{min}$ and $G_{max}$ are the minimum and maximum Gaussian curvatures corresponding to spherical plates of radius $r=26$ and $r=10$, respectively. The red dotted line is the minimum value attained in the set of numerical simulations for bubbles between concentric spheres, while the orange dotted line is the maximum value. Similarly, the maximum (orange dotted line) and minimum (red dotted line) values of relative plate separation $d^{*}=d\sqrt{|G(q)|}$ are shown in Fig \ref{gcurvature}c.

\subsubsection{Results}

\begin{figure}
\begin{center}
\includegraphics[width=1.0\columnwidth ]{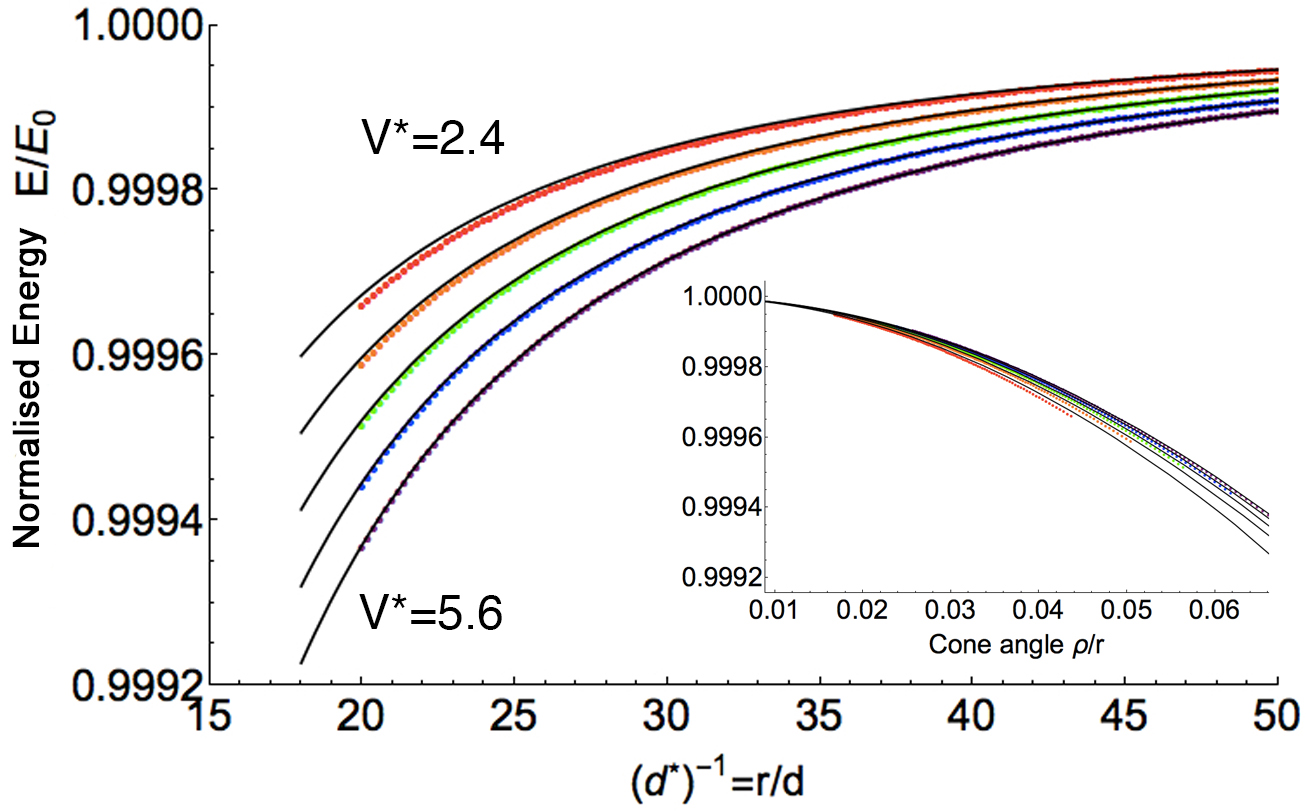}
\caption{Bubbles between spherical plates: effect of varying the volume and holding the plate separation constant. The graph shows a plot of the ratio $E/E_0$ as a function of the dimensionless radius $r/d$ (i.e. the inverse of the relative plate separation $d^{*}$), where $r$ is the radius of the inner plate and $d$ is the plate separation. Dots correspond to values from simulations while black lines are the expected energy of the bubble as given by Eq. ($\!\!$~\ref{eq:analytical1}). The inset shows the bubble energies plotted in terms of the (dimensionless) cone-angle.}
\label{spheres}
\end{center}
\end{figure}

In Fig \ref{spheres} we compare the energy of the simulated bubble with Eq. ($\!\!$~\ref{eq:scenergy}). Our normalised surface energy is given in terms of the expected surface energy $E_0$ of a bubble between flat plates: 
\begin{equation}
\frac{E}{E_0}
=
\frac{2}{\sigma_2}
\frac{\rho d}
{
r^2
}
\sqrt{
1
-
\left(
1
-
\frac{3V}{2\pi\sigma_3r^3}
\right)^2,
\label{eq:analytical1}
}
\end{equation}
which is obtained by combining Eq. ($\!\!$~\ref{eq:scenergy}) and Eq. ($\!\!$~\ref{eq:fenergy}). 

In Fig \ref{spheres} the horizontal axis shows the effect of an increasing inner shell radius. We can make the radius dimensionless by comparing it against the plate separation (since $d$ is constant in all simulations) and plot the normalised energies in terms of the dimensionless radius $r/d$ (i.e. the inverse of the relative plate separation $d^{*}$). An alternative representation is shown in the inset where the normalised energy is plotted against the cone angle. 

We conclude that when the bubble is bounded by plates of high (positive) Gaussian curvature the surface energy of the bubble is lowered. Or equivalently, increasing the cone angle lowers the energy of a bubble while a vanishing cone angle (i.e. a cylindrical bubble) has the largest possible energy.

\subsection{Concentric Tori}

We now consider the case of a bubble confined between two concentric (ring) tori. Here, the Gaussian curvature varies over the surface of the torus. 

\subsubsection{Plate Geometry}

\begin{figure}
\begin{center}
\includegraphics[width=0.75\columnwidth ]{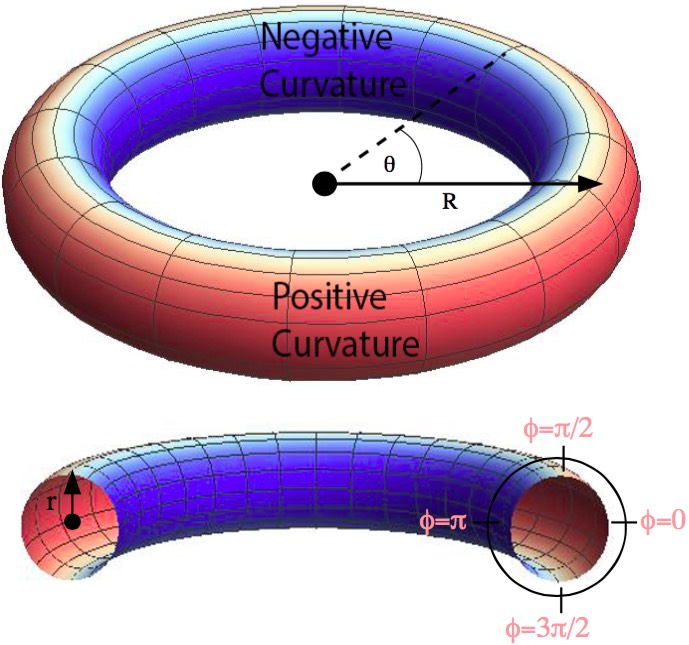}
\caption{Torus of radius $R$ and tube radius $r$ described by the parametric equation $( R-\sqrt{x^2 + y^2 } )^2 + z^2 - r^2 = 0$. The regions with maximum (positive) curvature are coloured red while regions of minimum (negative) curvature are coloured blue. The Gaussian curvature is zero between these two regions at $\phi=\pi/2$ and $\phi=3\pi/2$}
\label{torus_geometry}
\end{center}
\end{figure}

In the angular coordinates $(\theta, \phi)$ with $0\leq \theta \leq 2\pi, 0\leq \phi \leq 2\pi$ the parametrisation
\begin{eqnarray}
x&=&(R+r\cos(\phi))\cos(\theta), \nonumber\\
y&=&(R+r\cos(\phi))\sin(\theta), \nonumber\\
z&=&r\sin(\phi) \nonumber
\end{eqnarray}
defines a torus as the locus of points $(x,y,z)$ that satisfy the equation $\Phi(x,y,z,R)=r$
where,
\[
\Phi(x,y,z,R)=\sqrt{( R-\sqrt{x^2 + y^2 } )^2 + z^2}.
\]
The centre line of the torus is a circle of radius $R$ in the $x,y$-plane centred on the origin, and the toroidal tube itself has a radius $r$ -- see Fig \ref{torus_geometry}. This defines the inner toroidal plate while the outer plate is described by $\Phi(x,y,z,R)=r+d$, where $d$ is again the gap width. 

The Gaussian curvature on the inner toroidal plate is given by,
\begin{equation}
G(q)= G(\phi)=\frac{\cos(\phi)}{r(R+r\cos(\phi))},
\label{eq:Gtorus}
\end{equation}
so that on the outside of the torus it is a maximal (positive) value while on the inside it is a minimal (negative) value, as illustrated by the colour map in Fig \ref{torus_geometry}.

\subsubsection{Simulation Parameters}

We consider a series of small localised bubbles, beginning with $V=0.3$ and increasing in steps of $0.1$ up to $V=0.7$. We set $d=0.5$, $r=4.0$ and $R=20.0$, where the last two quantities are chosen to give a Gaussian curvature over the torus that is comparable to that of the spherical case, as shown in Fig \ref{gcurvature}a by the solid blue line.

Since the plate separation $d$ and the bubble volume $V$ have the same values as used in the simulations for bubbles between concentric spheres, the dimensionless volume of the bubbles is again in the range $2.4\leq V^*\leq5.6$. Also the geodesic radius of the bubbles is between the same limits, i.e.
\[
\rho_{min}=0.4370 \leq \rho \leq \rho_{max}=0.6676.
\]
These values are sufficiently small to satisfy the condition Eq. ($\!\!$~\ref{eq:condition}) everywhere. 

The cone angle and the relative plate separation, Eq. ($\!\!$~\ref{eq:scale}), are functions of the Gaussian curvature Eq. ($\!\!$~\ref{eq:Gtorus}) and as such depend on the azimuthal angle $\phi$. Both quantities are plotted in Fig \ref{gcurvature} and compared against the case for bubbles between spherical plates. The variation in these quantities -- in both the spherical and toroidal case -- are of the same magnitude. Consequentially, the variation in the surface energy in the two problems is also of the same magnitude, see Fig \ref{spheres} and Fig \ref{tori_vary_phi_1} (details of the toroidal simulations follow below). 

For the toroidal simulations, the maximum (purple solid line) and minimum (blue solid line) cone angle obtained with these simulation parameters is plotted in Fig \ref{gcurvature}a. The value of the relative plate separation is plotted in Fig \ref{gcurvature}b (blue line). Both of these values are significantly smaller than unity, which provides confidence that the cylindrical bubble model will be a valid approximation to the numerical results for the geometries investigated here.

\subsubsection{Results}

\begin{figure}
\begin{center}
\includegraphics[width=0.9\columnwidth ]{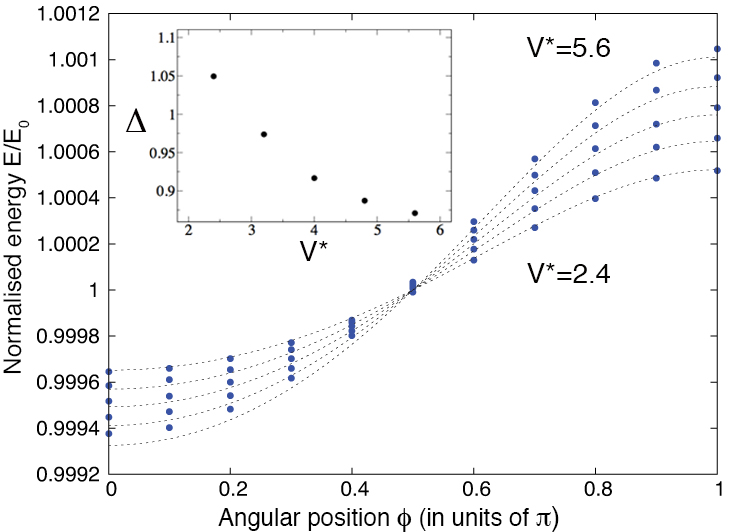}
\caption{
Bubbles between toroidal plates. The graph shows the energy of the bubble as a function of the local Gaussian curvature, which depends on the angle $\phi$. Simulations are conducted for a range of bubble volumes $V$ (in the plot we indicate bubble volumes in terms of the dimensionless volume $V^{*}=V/d^3$). Here we plot normalised energy $E/E_0$ as a function of the angle $\phi$. Blue dots correspond to values from simulations. Dotted lines are fits of the form given by Eq. ($\!\!$~\ref{eq:Tfit}). The correction $\Delta$ is found to be close to unity for small bubble volumes, indicating that for small bubbles the model works well (inset).}
\label{tori_vary_phi_1}
\end{center}
\end{figure}

The energy of the bubble is minimised, as described above. Once it is sufficiently converged we check the location of the bubble and confirm that the minimisation process has not displaced the bubble from its initial starting point (by computing the centre-of-mass of the vertices that comprise the mesh). As the bubble remains pinned during process 
we then compute the energy of the bubble as a function of its position (and therefore the local curvature of the bounding plates).

Since the Gaussian curvature of the torus does not depend on the $\theta$ coordinate we restrict ourselves to $\theta=0$ and vary the $\phi$  coordinate of the centre of the bubbl in discrete steps over the range $0 \leq \phi\leq \pi$. We restrict ourselves to the half range $0\leq\phi\leq \pi$ since by symmetry the Gaussian curvature of the torus is the same on the other side. 

In Fig \ref{tori_vary_phi_1} we plot the energy of the bubble in terms of the ratio $E(\phi)/E_{0}$ and compare it to a fit of the form
\begin{equation}
\frac{E(\phi)}{E_0}=1-\Delta\frac{E_1}{E_0}\frac{\cos(\phi)}{4(20+4\cos(\phi))},
\label{eq:Tfit}
\end{equation}
which is obtained by combining Eq. ($\!\!$~\ref{eq:approxE}) with Eq. ($\!\!$~\ref{eq:Gtorus})  and setting $R=20$, $r=4$. Here, $\Delta$ is a leading order correction, when $\Delta=1$ the model works perfectly while deviations indicate a break down of the model. As predicted by the model in the absence of curvature, when $\phi=\pi/2$, the energy of the bubble is given by $E_0$. The energy is at  a maximum when $\phi=\pi$ and at a minimum at $\phi=0$. 

There is generally good agreement between theory and numerical results, especially for small bubbles for which $\Delta$ is close to unity (see inset to Fig \ref{tori_vary_phi_1}). As the bubble volume increases this is no longer true, resulting in a gradual breakdown of the model. This is an indication that features neglected in our simple model (such as normal incidence, bubble curvature and variation in the Gaussian curvature over the bounding plates) become increasingly important for larger bubbles.

\subsection{Action of the geometric potential on bubbles}

\begin{figure}
\begin{center}
\includegraphics[width=0.9\columnwidth ]{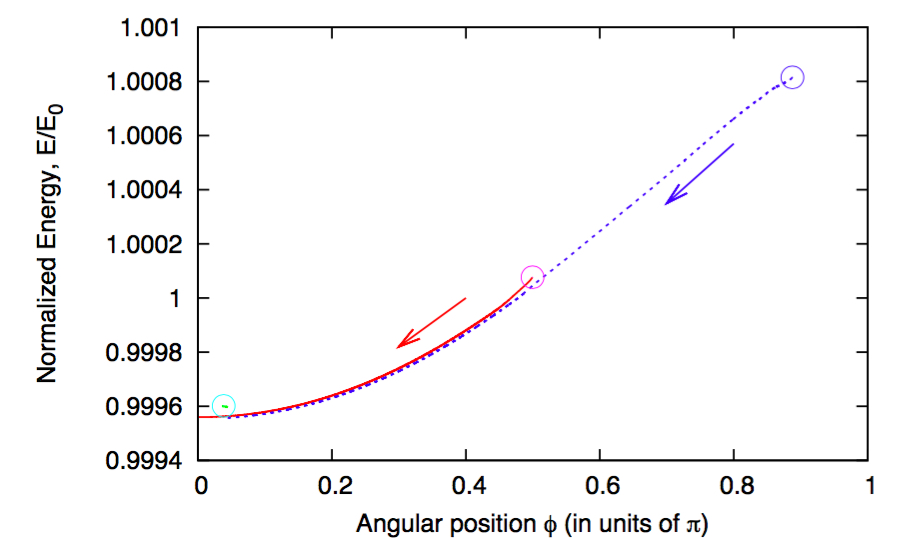}
\caption{Bubbles between toroidal plates move from regions of minimum (negative) curvature to regions of maximum (positive) curvature. The use of second derivative (Hessian of energy) information allows the bubble to move and further reduce its energy. Also see accompanying movie \cite{movie}.}
\label{torushessian}
\end{center}
\end{figure}

Even for modestly size bubbles ($V$ = 0.5, $d$=0.5) trapped between toroidal plates there is a significant difference in bubble energy as compared to whether the bubble is located on the outer or inner side of the torus. Fig \ref{torushessian} shows that this reduction in energy can drive a bubble from the region of negative curvature to the region of positive curvature. The motion is fastest around $\phi = \pi/2$, where the gradient $dE/d\phi$ is greatest (see Fig. 9), and a bubble at $\phi = \pi$ may be in a metastable state and not move.

The Surface Evolver simulations in Fig \ref{torushessian} were conducted with the same torus as those in \S IV.B, but the simulations are run for longer using second derivative information (Hessian of energy) to encourage bubble movement. We placed three bubbles between the toroidal shells, on the outside ($\phi=0$), inside ($\phi=\pi$) and top ($\phi=\pi/2$), and record their energy as they move around to the outside. The bubble on the inside starts a small distance from $\phi = \pi$ to avoid the region of metastability there. The bubbles usually stay close to the line $\theta = 0$, but they are not constrained to do so and we find that sometimes they do deviate from this shortest route to the outside of the torus (see accompanying movie \cite{movie}).

The motion of the bubbles is illustrated in the accompanying movie \cite{movie}, which shows three bubbles coloured red, blue and green. The blue bubble is initially located on the inner side of the torus, i.e. the region of minimum (negative) curvature, the red bubble is located on the outer side of the torus, i.e. the region of maximum (positive) curvature, while the green bubble is located at the flat point of the torus. The effect of the plate curvature is to drive the bubbles towards the outer side of the torus in order to minimise their energy.    

This effect can be thought of as a geometric potential (due to the shape of the bounding plates) which drives bubbles towards regions of maximum curvature. We anticipate that the magnitude of this potential will be greater for larger bubbles, this can be seen from the simple analytical model developed above. The change in energy due to plate curvature is given by
\[
\Delta E^{G}=E-E_0=-E_1G\propto -\frac{V^{3/2}}{d^{1/2}}G.
\]
Consider a cylindrical bubble of height $d$ and radius $\rho$. The cylinder has a volume $V=A d$, where the area of the circular base is given by $A=\pi\rho^2$. Thus for a small cylindrical bubble we can write the above result as
\[
\Delta E^{G}\propto -dA^{3/2}G,
\]
where $A$ is the apparent area of the bubble, i.e. the area of the bubble when it is projected onto one of the two bounding plates. Thus, provided the plate separation $d$ is constant, we expect the effect due to plate curvature on a small bubble to scale with the contact area between the bubble and the bounding plate. In a future publication we hope to demonstrate this; preliminary simulations suggest that the effect continues to magnify for bubbles that are well outside the range of our analytical model. 

\section{Perturbation to plate separation}
\label{sec:pert}

The above results demonstrate that bubbles confined between parallel curved plates can be driven from regions of minimal (negative) to maximal (positive) curvature. The examples we have considered include bubbles between concentric spheres and concentric tori. However, in any actual experiment that seeks to realise these results it is likely that the two bounding surfaces may not be exactly concentric (parallel). As such it is important to consider the relative importance of a variation in the plate separation as compared to the influence of plate curvature.

Suppose that we let  $d\rightarrow d+\delta \! d$ where $\delta \! d$ is a small change in the plate separation $d$. Then the surface energy of a cylindrical bubble is
\begin{eqnarray}
E\rightarrow E^{\delta \!d}
&=&
2\gamma \sqrt{\pi V (d+\delta\! d)}
\nonumber
\\
&=&E_0\left(1+\frac{\delta\! d}{d} \right)^{1/2}.
\nonumber
\end{eqnarray}
Assuming that $\delta\! d \ll d$ and expanding to first order we have
\begin{equation}
E^{\delta\! d}
\approx
E_0\left(1+\frac{1}{2}\frac{\delta\! d}{d}\right).
\nonumber
\end{equation}
Thus the change in surface energy of a bubble is
\begin{equation}
\Delta E^{\delta\! d}= E^{\delta\! d}-E_0=\frac{E_0}{2}\frac{\delta\! d}{d}.
\label{eq:diff1}
\end{equation}
Similarly, from Eq. ($\!\!$~\ref{eq:approxE}) we have
\begin{equation}
\Delta E^{G}=E-E_0=-E_1 G,
\label{eq:diff2}
\end{equation}
which is the difference in energy for a cylindrical bubble bounded by flat plates and curved plates. Taking the ratio of Eq. ($\!\!$~\ref{eq:diff1}) and Eq. ($\!\!$~\ref{eq:diff2}) gives,
\begin{eqnarray}
\frac{\Delta E^{\delta\! d}}{\Delta E^{G}}
\nonumber
&=&
-\frac{1}{2}\frac{E_0}{E_1}\frac{\delta\! d}{d}\frac{1}{G}\\
&=&
-
\frac{3\pi \delta\! d}{GV}.
\label{eq:Eratios}
\end{eqnarray}
Thus provided the amplitude of the perturbation in the plate separation obeys the condition
\[
|\delta\! d| \ll \left| \frac{GV}{3\pi} \right|,
\]
then the effects due to plate curvature will dominate. A further consequence of Eq. ($\!\!$~\ref{eq:Eratios}) is that for a bubble of a given volume the change in energy due to the curvature of the plate can be compensated for by an equivalent change in the local plate separation.

In the case of a sphere of radius $r=12$ and a bubble of volume $V=0.5$, we have $|GV/3\pi|=0.00036$. Expressed as a fraction of the sphere radius the magnitude of the perturbation must be significantly less than $\delta \!d/r=0.00003$ (i.e. about $0.003\%$). This implies that for the effect to be significant for small bubbles the bounding plates must be arranged with considerable precision. However, for larger bubbles (or equivalently if the bounding surfaces have a large Gaussian curvature) the effect can be significant enough to drive a bubble from a region of low (negative) curvature to regions of high (positive) curvature.

\section{Discussion}
\label{sec:disc}

Through a combination of analytical and numerical methods we have investigated the properties of a single area-minimising bubble confined between two parallel curved plates. Our work demonstrates that the surface energy of the bubble depends on the local curvature of the plates. This energy difference can be large enough to drive bubbles from regions of negative Gaussian curvature to regions of positive Gaussian curvature. 

The motion of a bubble along a path of increasing Gaussian curvature (from negative to positive) is driven by the decrease of surface energy. We can adopt the perspective that the curved plates impose a `geometric potential' that drives the bubble. This potential contributes to the energy landscape of the system, in addition to gravity or other sources such as variations in plate separation. It would be interesting to determine whether even a small frictional resistance would stop this movement, or to what extent the presence of a wetting film on the surfaces of the tori would allow the movement to be observed.

We consider the perspective of a geometric potential as useful in particular when addressing cellular foams composed of many bubbles which may cover large areas of varying curvatures. The statistical mechanics of this cell collective -- that for example determines its degree of disorder or crystallinity -- will then take place against the background of this geometric potential. Various of the commonly addressed questions in foam physics can be revisited in the context of single- or multi-bubble systems confined between curved plates (``warped Hele Shaw cells''), as opposed to the usual case of flat confinement. These phenomena include diffusive coarsening (note early work on this phenomenon in spherical systems \cite{roth2012coarsening, senden})  and the role of topological defects in relaxing stress.

We see the soft deformable soap froth as an interesting alternative to the more commonly studied particulate packing or assembly problems \cite{bausch2003grain, irvine2010pleats, dotera2012hard}. The present system is richer, as it experiences a more immediate, more local, force or potential than the particulate systems, where defect formation results when the packing problems reaches large enough length-scales for the topological Gauss-Bonnet theorem to become significant. Understanding the interplay between the geometric potential acting on bubbles (as described here) and the role of topological defects in crystalline arrangements on curved surfaces will be the subject of future investigations. 

A further possibility is that of bubbles (and foams) confined between curved surfaces with vanishing Gaussian curvature, such as between concentric cylinders. In such cases we expect any variation in the energy of the bubble to depend entirely on the local mean curvature. It would be useful to compare such mean curvature effects against the present model.
\newline 

\section{appendix}
Here we list the immediate higher order terms for the circumference and area of a Geodesic circle, located at a point $q$ on some surface with a local Gaussian curvature $G(q)$. The circumference $C(q)$ is given by,
\begin{equation}
C(q)=2\pi \rho - \frac{\pi \rho^3 }{3}G(q)+ \frac{\pi \rho^5 }{60}[G(q)]^2 + \mathcal{O}(\rho^5)
\label{eq:theorem}
\end{equation}
while the area $A(q)$ of the geodesic circle is,
\begin{equation}
A(q)=\pi \rho^2 - \frac{\pi \rho^4 }{12}G(q) + \frac{\pi \rho^6 }{360}[G(q)]^2 +
\mathcal{O}(\rho^8).
\label{eq:Atheorem}
\end{equation}

\section{Acknowledgements} We thank Ken Brakke and Andy Kraynik for useful discussions and for help with the implementation of the Surface Evolver model. We also acknowledge useful discussions with Denis Weaire. SJC acknowledges funding from EPSRC (EP/N002326/1). A. M. acknowledges support from the Aberystwyth University Research Fund. AM and GS-T acknowledge funding from ``Geometry and Physics of Spatial Random Systems'' under Grant No. SCHR-1148/3-2. 

\bibliographystyle{nonspacebib}

\end{document}